\documentclass[11pt,a4paper]{article}
\usepackage[margin=1.04in]{geometry}
\usepackage[T1]{fontenc}
\usepackage[utf8]{inputenc}
\usepackage{lmodern,microtype,amsmath,amssymb,booktabs,threeparttable,tabularx,array}
\usepackage[round,authoryear]{natbib}
\usepackage[hidelinks]{hyperref}
\usepackage{xurl,graphicx,caption,setspace,float,enumitem}
\newcommand{\ind}{\mathbf{1}}
\title{\bfseries The Private Production of Safety\\[0.27em]\large Outside Options and the Burden of Protection}
\author{Elena Pisanelli\\[0.2em]\normalsize University of Bergamo, Department of Economics}
\date{}
\begin{document}\maketitle

\begin{abstract}
\noindent Protective interventions are typically evaluated by their ability to prevent harm. Yet protection also raises a distinct question: who bears the burden of adjusting to violence? I examine this allocation of protective adjustment using four randomized interventions in the United States. An offer of permanent housing assistance increases separation from a baseline partner combined with stable independent housing by 12.52 percentage points. Assignment to active police-response protocols reduces observed victim-only departures and increases suspect-only departures. Assignment to an emergency protection order increases documented directions for suspects to leave. These findings reveal two margins of the same allocation problem: housing assistance expands victims’ ability to establish an independent residence, while enforcement shifts displacement obligations toward suspects. Protective policies thus shape not only the opportunities for safety, but also the distribution of the costs required to achieve it.
\end{abstract}
\clearpage
\section{Introduction}\label{sec:intro}
The effectiveness of a protective intervention is commonly assessed by asking whether it prevents harm. For people threatened by violence, another consequence matters: who must alter their circumstances in order to be safe. A person may leave a shared home, establish a separate residence, or prepare for a subsequent incident. A legal order may instead require the suspected aggressor to leave. Public assistance can also change the terms on which leaving is possible. I study how policy affects this \emph{incidence of protective adjustment}: the identity of the person who acts and the conditions under which that action takes place.

This distinction has a straightforward economic foundation. People respond to the alternatives available to them and to the obligations that institutions impose on others. A housing subsidy improves the feasibility of living independently without requiring any relationship to end. An order issued during a police encounter may redirect the immediate adjustment from the person seeking protection to the person against whom the order is issued. Both changes can matter even when a single aggregate outcome captures little of their effect. A count of separations says nothing about the housing conditions that follow them; an indicator for whether anyone left a scene says nothing about which person moved.

The distinction also matters for interpreting the term \emph{burden of protection}. I use it to refer to the actions and disruptions a person may have to undertake to obtain safety, not merely to expenditures or lost earnings. It would be misleading, however, to treat every departure or precaution as a cost. A person who separates because a better alternative is now available exercises a choice. A person who must abandon a residence to escape danger faces a different constraint. The empirical objective is to distinguish these margins before attaching a welfare interpretation to any one of them.

I bring new evidence to this question from randomized housing assistance and police-response experiments in the United States. The first setting is the Family Options Study, which offered different forms of housing assistance to families recruited from emergency shelters. The experiment changed access to independent living in a population broader than people previously identified as victims of intimate-partner violence. The original evaluation found that the offer of potentially permanent assistance increased the likelihood of separation from a partner present at the start of the study \citep{gubits2016}. I take a step forward and examine the residential circumstances in which those separations occur by linking relationship and housing records for the same families. The policy increases the probability of \emph{separation accompanied by stable independent housing} by 12.52 percentage points at 37 months among families with both outcomes observed. It also reduces by 19.07 points the probability of remaining partnered while lacking stable independent housing. 

The second setting is a randomized police-response experiment in Colorado Springs. At a violent encounter, an enforcement policy can change which person is directed to leave and which person actually departs. I reconstruct the four mutually exclusive scene states: the victim leaves alone, the suspect leaves alone, both leave, or neither leaves. Across 1,087 encounters with both departure reports, the active response packages reduce victim-only departures by 11.97 percentage points and increase suspect-only departures by 9.43 points relative to ordinary response. There is no statistically significant difference in the probability that someone departs. The difference appears in \emph{who moves}, not in the overall frequency of movement. Separately, among all 639 assignments in the order-only and ordinary-response arms, assignment to an emergency protection order increases documented directions for suspects to leave by 55.87 points. The administrative order and the observed departure identify different stages of enforcement.

The analysis builds on economic models of household bargaining and abusive relationships. Early work links domestic violence to household resources and the incentives governing partners' decisions \citep{tauchen1991,farmer1997,bloch2002}. Subsequent evidence studies how women's labor-market opportunities, transfers, and control over resources affect violence and autonomy \citep{aizer2010,bobonis2013,eswaran2011}; \citet{adams2024} emphasize how economic suppression itself can undermine the ability to exit. Changes in divorce law also alter the terms of relationship dissolution and family distress \citep{dee2003,stevenson2006}. These studies motivate the role of outside options, but the outcomes of interest here are distinct: whether a relationship change is accompanied by a sustainable residence, and which person undertakes the immediate adjustment when police intervene.

A second connection concerns the technology and consequences of protection. Models of self-protection and private precautions explain why an individual's actions need not coincide with the socially preferred response to risk \citep{ehrlich1972,shavell1991}. Housing instability can compound the circumstances facing vulnerable families \citep{browne1997,desmond2015}. The original Family Options Study evaluates housing-assistance offers \citep{gubits2016}, and recent work uses that experiment to estimate effects of housing access on partner violence \citep{winichakul2026}. Separately, randomized arrest studies and research on enforcement laws ask how police responses affect repeat violence and deterrence \citep{sherman1984,berk1992,pate1992,sherman1992,maxwell2002,iyengar2009}. Work on protective resources also distinguishes reductions in exposure from possible retaliation \citep{dugan2003}. I examine a complementary, directly observed outcome: the \emph{distribution of protective adjustment} across people and housing circumstances, without equating changes in those actions with changes in violence.

I make three empirical contributions. First, building on the housing experiment and research on outside options, I treat relationship separation and residential stability as a joint outcome rather than interpreting separation alone as evidence of a viable exit \citep{gubits2016,winichakul2026}. Second, I distinguish a police-recorded legal direction from realized movement and decompose scene departures by person, an outcome dimension different from the recurrence and deterrence measures emphasized in the police-response literature \citep{berk1992,maxwell2002}. Third, I examine two additional randomized interventions---advocacy in Harris County and detention in Milwaukee---to distinguish support for private safety preparations from legally directed displacement and subsequent safety-motivated relocation \citep{mcfarlane2002,mcfarlane2004,sherman1992}. These experiments identify distinct policy margins; considered together, they reveal how public interventions reshape the constraints, opportunities, and obligations that determine who bears the burden of achieving safety.

The paper proceeds in the order of the empirical designs. Section~\ref{sec:data} describes the programs, source data, and observation samples. Section~\ref{sec:framework} defines protective adjustment and the assignment-based estimands. Sections~\ref{sec:hud} and \ref{sec:colorado} present the main results, followed by the advocacy and detention evidence. Section~\ref{sec:robustness} reports specification and measurement checks. Section~\ref{sec:discussion} discusses implications for the design and evaluation of legal protection.

\section{Institutional settings and data}\label{sec:data}
\subsection{Housing assistance and relationship outcomes}
The Family Options Study randomized 2,282 families recruited from emergency shelters in twelve U.S. communities to offers of different housing programs \citep{gubits2016}. The principal alternatives were a potentially permanent housing subsidy, project-based transitional housing, community-based rapid rehousing, and usual care. They differed in the duration and organization of assistance. A potentially permanent subsidy made long-term rental support available; transitional housing combined accommodation with services for a limited period; rapid rehousing provided shorter-term assistance. Usual care consisted of the services otherwise available to families in the local community.

Randomization took place among the alternatives available to a family at the time of assignment. It is therefore important not to treat the four program names as if every family faced the same four-way choice. I analyze the original comparison between an offer of a permanent subsidy and usual care and retain the information about program availability recorded before randomization. Assignment is to the \emph{offer}, not to eventual use of assistance. This preserves the policy question of how access to an outside option changes subsequent circumstances, rather than comparing families who selected into different housing arrangements.

The sample used in this paper is also distinctive in relation to the question of violence. Families were recruited because they were living in emergency shelters, not because a partner had been identified as violent. Prior evidence documents how housing insecurity and intimate violence can overlap among economically vulnerable women \citep{browne1997}, but this sample cannot be treated as a cohort of verified abuse survivors. The policy could affect the conditions for a relationship decision among both people with and without previously reported victimization. This is important for interpreting separations: the experiment concerns the value of an outside option, and I do not assume that each observed separation is an escape from abuse.

I link the original public-use baseline and adult follow-up files by their compatible anonymized identifiers. Among families eligible for the permanent-subsidy versus usual-care comparison, 313 report a spouse or partner at baseline: 153 assigned to the subsidy and 160 to usual care. At 37 months I observe both the relationship outcome and the components of residential stability for 124 and 116 families, respectively, a total of 240. The corresponding 20-month linked sample contains 253 families. I retain original assignment counts alongside the observed-outcome denominators throughout the paper.

I classify a family as having an \emph{independent, stable home} if the follow-up records indicate residence in its own home, no episode of homelessness or doubled-up accommodation during the preceding six months, and no change of residential location over that interval. The measure is intentionally stricter than residence in one's own home on the interview date. It records the durability of the residential alternative, the outcome most directly related to the burden of maintaining a separate household. I cross-classify this indicator with separation from the partner present at baseline. The resulting four states are exhaustive for families with complete linked outcomes.

The housing measure has two features important for interpretation. First, the initial partner is identified before the offer, so the definition of the population does not depend on whether the subsequent relationship dissolved. Second, the residential indicator covers the six months preceding the follow-up interview, rather than the survey day alone. A family occupying its own unit only briefly is distinguishable from a family that has maintained an independent residence without another move. By crossing the two variables, the analysis records not just whether an alternative to the initial relationship was exercised but also the observable durability of that residential alternative.

The public-use follow-up survey is not a balanced panel in which every baseline participant answers each item at both dates. The number contributing to the joint state is therefore given separately at 20 and 37 months. The estimates preserve the original assignment and use the available comparison-specific nonresponse weights. That approach is consistent with studying the effect of an \emph{offer} on the joint outcomes observed at a specified follow-up, rather than conditioning on the eventual receipt of the subsidy.

\subsection{Police encounters and immediate departures}
The Colorado Springs experiment assigned eligible spouse-assault encounters to four response packages: arrest combined with an emergency protection order, counseling combined with an order, an order alone, and ordinary restoration of order \citep{icpsr9982,berk1992}. The assignment unit was the encounter. The treatments were not a factorial experiment in which arrest, counseling, and the order were independently switched on and off. I consequently estimate effects of the response \emph{packages} and preserve the original arms in the analysis.

The original archive contains two kinds of record needed for my question. Police administrative entries document the direction that officers recorded in connection with a protection order. Scene reports indicate whether the victim and the suspected aggressor left. An order directed at a suspect is legally meaningful even if an accompanying departure report is absent; conversely, a departure is an observed action rather than proof of compliance with a particular legal order. I analyze these outcomes separately.

The source reconstruction yields 1,314 first-valid eligible assignments from the selected 1988--1989 cohort, distributed as 329 arrest-plus-order, 346 counseling-plus-order, 337 order-only, and 302 ordinary-response encounters. Both scene-departure indicators are observed in 1,087 encounters (82.7 percent). The proportions with both outcomes observed are 84.2, 81.5, 81.0, and 84.4 percent in the four arms. The full-assignment administrative contrast between the order-only and ordinary-response arms uses all 639 assignments in those arms. For later safety reports, I keep police-known incidents and victim interviews separate because their observation windows and reporting processes differ.

The distinction between directions and departures is partly a distinction between the object of law and the object of measurement. An assigned emergency order is a police-response protocol; an officer's recorded direction is a documented implementation outcome; a departure is a behavior observed afterward. Each has a different denominator and responds to a different question. The scene reports are available in a larger number of cases than a narrowly defined subset of physical-violence follow-up interviews, but in fewer cases than the original assignment record. I retain the original four-way randomization because pooling the three active packages is analytically convenient but does not make them identical interventions.

The 1,314-case cohort is not the universe of every record contained in the source archive. The archive has 1,660 source records, from which I reconstruct the eligible first-valid assignment cohort for 1988--1989. This construction prevents repeated or ineligible records from being counted as independent randomized assignments. 

\subsection{Advocacy and detention}
Two additional randomized experiments help interpret the kinds of action that institutions may change. In Harris County, 150 women were assigned in equal numbers to an advocacy intervention or comparison \citep{icpsr3542,mcfarlane2002}. Advocacy supplied assistance with safety planning. The original records contain an indicator for whether a protection order was obtained and repeated reports on precautions such as keeping emergency cash, having spare keys, and preparing an emergency bag. I use the fraction of valid follow-up waves in which each preparation was reported, controlling for its baseline value. A response coded ``not applicable'' remains missing rather than being classified as an absence of action.

The Milwaukee experiment instead randomized 1,200 incidents among overnight arrest, short-release arrest, and advice \citep{icpsr9966,sherman1992}. I use the item asking whether the respondent subsequently stayed with friends, relatives, or at a shelter to avoid being hurt, as well as later incidents recorded by police. The former directly describes a later safety-motivated adjustment; the latter records administrative recurrence. The temporary-relocation item is available for 919 assignments. These designs complement Colorado because they measure preparation and later temporary relocation rather than immediate departure at the scene.

The four settings differ in their assignment unit, follow-up window, and principal behavioral outcome. The housing study concerns families at risk of housing instability; Colorado concerns the parties at a police encounter; Harris concerns women receiving advocacy; and Milwaukee concerns follow-up after a randomly assigned police sanction. Their variation is useful because it separates types of protective action that would otherwise be difficult to distinguish.

\section{Economic interpretation and empirical strategy}\label{sec:framework}
The experiments are included for a specific reason. The principal housing and police comparisons identify a durable residential alternative and the identity of the person leaving an encounter, respectively. Neither outcome on its own captures every kind of adjustment associated with safety. Advocacy allows an individual to prepare without relocating, and Milwaukee measures a later move explicitly undertaken to avoid harm. These settings help distinguish \emph{capacity to act}, \emph{a legal direction to act}, and \emph{a reported action taken for protection}. The paper estimates each effect within its own randomized comparison.

\subsection{Outside options and the circumstances of separation}
A household's decision to remain together depends partly on the alternatives available to each person. Access to an independent residence can change this comparison even where violence has never been reported. To make that margin explicit, consider a person who compares the expected value of living apart with the expected value of remaining with a partner. A simplified decision rule is

\begin{equation}
 u_i^{\mathrm{out}}(O_i)-d_i^{\mathrm{move}}\;\geq\;u_i^{\mathrm{stay}}-\pi_i h_i,
 \label{eq:exit}
\end{equation}

where $O_i$ denotes available residential alternatives, $d_i^{\mathrm{move}}$ the disruption of moving, and $\pi_i h_i$ the expected harm associated with remaining. This expression is a description of the choice margin, not a structural equation estimated in the paper. A subsidy can improve $u_i^{\mathrm{out}}(O_i)$ even when expected harm is low. It can also make independent living sustainable after separation. The first implication concerns the decision to separate; the second concerns the conditions under which separation takes place.

I measure those conditions directly. Let $R_i$ denote separation from the baseline partner, and let $H_i$ denote an independent, stable home. The four potential joint outcomes $(R_i(z),H_i(z))$ are $(1,1)$, $(1,0)$, $(0,1)$, and $(0,0)$. Write $p_{rh}(z)$ for the probability of state $(r,h)$ under assignment $z$. The policy effect on each state is $p_{rh}(1)-p_{rh}(0)$; the four effects sum to zero. Separate marginal effects on separation and housing stability can be obtained from these cells, but two marginals alone cannot recover their joint distribution. This is why a new cross-classified outcome is useful: it distinguishes \emph{separation with a durable outside option} from separation accompanied by continued residential instability.

Crucially, the joint state does not classify separations as voluntary or forced, and it does not identify residential stability as a mediator of the treatment effect. It describes an economically meaningful combination of observed outcomes after random assignment. In particular, I do not condition the separation regression on the family's post-treatment housing status; instead, I estimate the treatment effect on the joint state itself.

This formulation separates two margins that are often collapsed in policy evaluation. The first is the \emph{extensive margin of adjustment}: whether a change of residence or relationship occurs. The second is the \emph{incidence of adjustment}: the conditions of the change and the person undertaking it. The distinction is not limited to violent relationships. For a family offered rental assistance, the opportunity to establish a separate household may be valuable even when there is no documented violence at baseline. For a person at a police encounter, by contrast, whether a suspect receives a direction to leave may matter immediately, regardless of longer-run household choices.

A second implication concerns welfare interpretation. An outside option does not operate like a compulsory move. Giving someone a subsidized housing opportunity can improve the feasible set whether or not the individual chooses to use it. An order compelling another person to leave operates through legal authority and creates a direction that can be documented separately from compliance. The joint-state and scene-departure measures are therefore best understood as empirical distributions generated by two different technologies of protection. The analysis establishes how their components respond to assignment without imposing an arbitrary ranking of every separation, departure, or precaution.

\subsection{The person who leaves during police intervention}
The immediate incidence of a police response can be expressed without assigning a numerical welfare cost to moving. Let $V_i$ be an indicator that the victim leaves the scene and $S_i$ an indicator that the suspected aggressor leaves. Define

\begin{equation}
Q_i=\ind\{V_i=1\ \mathrm{or}\ S_i=1\},\qquad
I_i=S_i-V_i=\ind\{S_i=1,V_i=0\}-\ind\{V_i=1,S_i=0\}.
\label{eq:incidence}
\end{equation}

The variable $Q_i$ indicates whether at least one person leaves, while $I_i$ distinguishes the direction of one-person departure. In particular, replacing a victim-only departure with a suspect-only departure raises $I_i$ by two units while leaving $Q_i$ unchanged. An increase in the signed index can therefore occur without any change in total departures. This is an accounting property of the outcomes, not a hypothesis about the effect of arrest or an assumption about whose welfare has increased.

The experiment also records a policy-relevant step before movement: whether officers document a direction requiring the suspect to leave. I treat that administrative indicator as a separate outcome. The distinction is central to an interpretation in law and economics. A directive allocates an institutional obligation; realized departure records behavioral adjustment. They can move together, but one is not a substitute for the other.

The signed index has an additional useful interpretation. Because $I_i$ equals suspect departure minus victim departure, its treatment contrast is the effect on suspect departures net of the effect on victim departures. When both people move, their contributions cancel in the signed index, but that state remains observable in the full four-state analysis. Reporting the index alongside its constituent states guards against a change driven by a mechanically aggregated measure rather than by who is observed adjusting. It also allows the analysis to compare the signed direction of movement with the probability of any movement without confusing the two.

\subsection{Estimation and interpretation}
For the Family Options Study I estimate the following linear probability model separately for each joint state and survey wave:

\begin{equation}
Y_{iw}=\alpha_w+\beta_w\,\mathrm{SUB}_i+\sum_{a=2}^{A}\gamma_{aw}\,\ind\{A_i=a\}+\varepsilon_{iw},
\label{eq:hudreg}
\end{equation}

Here $\mathrm{SUB}_i$ equals one for assignment to the permanent-subsidy offer, and $A_i$ records the set of program alternatives available before randomization. The coefficients on availability indicators account for the original comparison structure. I use the public-use comparison- and wave-specific response weights and calculate heteroskedasticity-consistent HC2 standard errors under the weighted transformed design. The reported sample consists of initially partnered families with both outcome components observed at the relevant follow-up. Unweighted models and models without availability adjustments show how much the result depends on those choices.

For Colorado, with the ordinary-response arm as the benchmark, the pooled response-package comparison is

\begin{equation}
Y_i=\alpha+\tau\,\ind\{Z_i\in\{1,2,3\}\}+u_i,
\label{eq:coreg}
\end{equation}

where the indicator identifies any of the three active assignments. I also estimate arm-specific contrasts against ordinary response and a specification absorbing study year. This is an analysis of assignment to a response package, not of the treatment an officer eventually chose to implement. When an outcome is incomplete, the table identifies its observed sample. For the police-recorded order outcome the order-only comparison uses the full assigned sample for those two arms; the movement outcomes use the 1,087 encounters with both indicators recorded. The design thus isolates policy assignment while maintaining the distinction between observable institutional action and observed behavior.

\section{Housing assistance and the conditions of separation}\label{sec:hud}
\subsection{Relationship decisions and residential alternatives}
The housing experiment changes the outside option of an initially sheltered family. The original evaluation found that the offer of potentially permanent assistance increased separation from a partner present at baseline by 13.4 percentage points at 37 months, with adjusted separation shares of 47.8 percent under the subsidy offer and 34.3 percent under usual care \citep{gubits2016}. The estimate concerns the broader group of families with a baseline partner, not only those reporting violence. It establishes that the offer affects relationship status; it does not indicate whether the people who separate subsequently live independently and stably.

That distinction motivates my new outcome. An initially sheltered family can report a separation and nevertheless move repeatedly among provisional arrangements. Alternatively, the same transition can coincide with continuous residence in its own home. The housing and relationship files make it possible to measure these states together. I focus on that joint distribution rather than treating the relationship decision alone as evidence that the outside option has become durable.

\subsection{The four joint states at 37 months}
Table~\ref{tab:hud-joint} reports the weighted, program-availability-adjusted assignment effects for the four exhaustive states. Among 240 families with linked relationship and residential outcomes, the offer of a permanent subsidy increases separation accompanied by a stable independent home by 12.52 percentage points. At the same time, the share reporting no separation and unstable housing falls by 19.07 points. The estimates for separation with unstable housing and for remaining partnered with stable housing are not statistically significant individually.

\begin{table}[htbp]
\centering
\caption{HUD: relationship separation and residential stability, measured jointly}
\label{tab:hud-joint}
\begin{threeparttable}
\small
\begin{tabular}{p{9.5cm}r}
\toprule
Status at 37 months & SUB--UC (pp) \\
\midrule

Separated; independent, stable housing
    & $+12.52^{**}$ \\
    & $(5.33)$ \\
\addlinespace[0.5em]

Separated; unstable housing
    & $-4.91$ \\
    & $(5.34)$ \\
\addlinespace[0.5em]

No recent separation; stable home
    & $+11.47^{*}$ \\
    & $(6.53)$ \\
\addlinespace[0.5em]

No recent separation; unstable housing
    & $-19.07^{***}$ \\
    & $(5.16)$ \\

\midrule

All stable housing (irrespective of separation)
    & $+23.99^{***}$ \\
    & $(6.35)$ \\

\bottomrule
\end{tabular}

\begin{tablenotes}
\footnotesize
\item \emph{Notes:}
Author's analysis of authentic HUD PUF1 linked microdata.
N=240 families with both relationship and housing outcomes observed,
out of 313 baseline-partnered families assigned in the original
SUB--UC eligible comparison.
SUB=permanent-subsidy offer; UC=usual care.
Weighted linear probability models adjust for pre-randomization
program availability. HC2 standard errors in parentheses.
$^{*}p<0.10$, $^{**}p<0.05$, $^{***}p<0.01$.
Stable housing requires one's own residence, no homelessness or
doubling up in the preceding six months, and one residential
location in that period.
The final row sums the first and third states and is not an
additional independent partition.
Separations are not restricted to violent partners.
\end{tablenotes}

\end{threeparttable}
\end{table}
The four coefficients sum to zero by construction. Their composition is informative: the increase in stable housing is not confined to families who separated, nor does the increase in separation necessarily describe entry into instability. Combining the two stable states, the subsidy raises independent, stable housing by 23.99 points. These effects reveal the residential circumstances associated with a relationship transition, rather than inferring those circumstances from aggregate housing or separation rates.

\begin{figure}[htbp]\centering
\includegraphics[width=.88\linewidth]{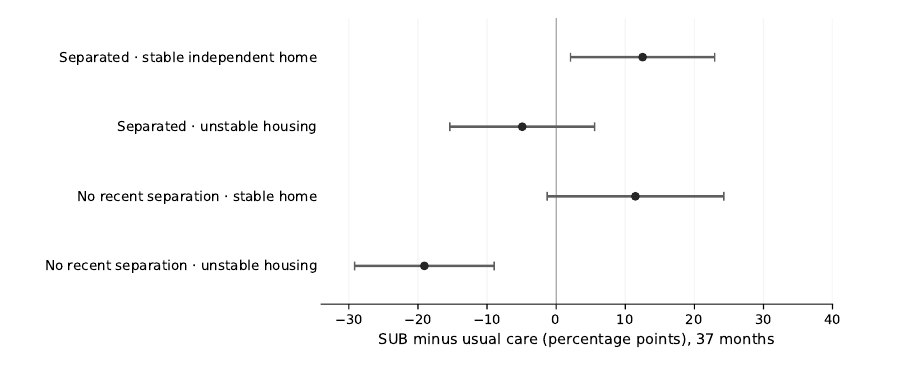}
\caption{Permanent subsidy versus usual care: four joint relationship--housing states at 37 months. Points are weighted, program-availability-adjusted assignment contrasts, with HC2 pointwise 95\% intervals among 240 respondents.}
\label{fig:hud-joint}\end{figure}

Figure~\ref{fig:hud-joint} shows the four joint-state estimates with their pointwise intervals. It is useful to compare the first and fourth states. The first records separation accompanied by a durable home; the fourth records continued partnership together with residential instability. Their opposite movements show that the policy changes both the composition of relationship outcomes and the quality of available housing. Because the four outcomes are exhaustive, I interpret them as a distribution and avoid treating their separate coefficients as four unrelated causal mechanisms.

The accounting decomposition also helps interpret the size of the effect. The 12.52-point increase in separation with a stable home accounts for just over half of the 23.99-point increase in stable housing across both relationship categories. The remainder, 11.47 points, comes from families who are not observed separating from the baseline partner. The latter individual-state estimate has an interval that includes zero; the decomposition is an identity among the four reported point estimates, not evidence that one relationship category mediates the other. This is precisely the information a separate separation indicator or a separate stability indicator would not provide: the subsidy changes residential conditions in more than one relationship state.

Another informative feature of Table~\ref{tab:hud-joint} is that the separated-and-unstable coefficient is negative rather than positive, although statistically indistinguishable from zero. The increase in separation with a stable home is therefore not mechanically accompanied by a detectable increase in separated families experiencing residential instability. This comparison addresses the character of the outside option, not the motives behind the original relationship decision. It is a direct description of the joint states produced after an offer of permanent assistance.

\subsection{Earlier follow-up and the interpretation of choice}
At 20 months, among 253 families with linked outcomes, the separation-and-stable-housing contrast is 5.32 percentage points  and is not statistically significant. The effect on stability irrespective of separation is 15.63 points. At 37 months the corresponding point estimates are 12.52 and 23.99 points. The two follow-up samples differ; these are therefore separate assignment comparisons at each wave rather than estimates of individual changes between surveys. The pattern is consistent with stable arrangements taking time to develop, but it does not rely on interpreting a cross-wave difference as a treatment effect on duration.

In the linked 37-month sample, the marginal separation contrast is 7.60 points, compared with the larger separation estimate in the original evaluation. The underlying analysis samples and weighting procedures are different. The distinctive result in this paper is the joint state, which records separation \emph{together with} residential continuity. It is relevant precisely because the experiment did not enroll only people known to face a violent partner. An improved outside option changes the opportunities for relationship choice more generally; the joint outcome shows what residential arrangements accompany those choices.

\section{Police response and the allocation of departure}\label{sec:colorado}
\subsection{From a binary movement indicator to four states}
At a police encounter, an indicator for whether somebody leaves combines substantively different outcomes. In the Colorado data I can observe the four exclusive states for cases with reports on both people. Under ordinary response, the victim alone leaves in 44 of 255 recorded encounters (17.25 percent), while the suspect alone leaves in 59 (23.14 percent). Under the three active response packages together, victim-only departures are reported in 44 of 832 encounters (5.29 percent) and suspect-only departures in 271 (32.57 percent). The remaining cases involve either both people leaving or neither leaving. Table~\ref{tab:colorado-states} reports all four states by original assignment.

\begin{table}[htbp]\centering
\caption{Colorado Springs: original assignment and departure states}\label{tab:colorado-states}
\begin{threeparttable}\small
\setlength{\tabcolsep}{3.8pt}\scriptsize\begin{tabular}{lrrrrrr}\toprule
Assigned protocol & Assigned & Pair observed & Victim only & Suspect only & Both & Neither\\\midrule
Arrest + EPO & 329 & 277 & 10 (3.6\%) & 93 (33.6\%) & 14 (5.1\%) & 160 (57.8\%)\\
Counseling + EPO & 346 & 282 & 15 (5.3\%) & 78 (27.7\%) & 19 (6.7\%) & 170 (60.3\%)\\
EPO alone & 337 & 273 & 19 (7.0\%) & 100 (36.6\%) & 16 (5.9\%) & 138 (50.5\%)\\
Ordinary response & 302 & 255 & 44 (17.3\%) & 59 (23.1\%) & 13 (5.1\%) & 139 (54.5\%)\\\midrule
Total & 1,314 & 1,087 & 88 & 330 & 62 & 607\\\bottomrule
\end{tabular}
\begin{tablenotes}\footnotesize\item \emph{Notes:} Public ICPSR 9982 first-valid assignments. Percentages use cases with both departure indicators observed in each original assigned arm. EPO=emergency protection order. Assigned protocols are bundled treatments, not a factorial randomization of separate components.\end{tablenotes}
\end{threeparttable}\end{table}

These frequencies suggest why aggregate movement can miss an important consequence of enforcement. The total probability of somebody departing combines the one-person states with the both-person state. It does not identify the person on whom the adjustment falls. The signed index in equation~\eqref{eq:incidence} preserves that information while treating a case in which both people depart as different from a case in which only one person does.

\subsection{The legal direction recorded by police}
I first examine a direct institutional outcome using the complete assigned comparison between an emergency protection order alone and ordinary restoration of order. The police records document a direction to the suspected aggressor to leave in 224 of the 337 order-only assignments (66.47 percent) and 32 of the 302 ordinary-response assignments (10.60 percent). The difference is 55.87 percentage points. Unlike the paired departure classification, this contrast uses all 639 assigned encounters in the two arms.

\begin{table}[htbp]
\centering
\caption{Colorado: assignment to an order and documented legal direction}
\label{tab:colorado-legal}
\begin{threeparttable}
\small
\begin{tabular}{p{6.5cm}rrr}
\toprule
Assigned response
    & Assigned
    & Recorded direction
    & Share (\%) \\
\midrule

Emergency protection order alone
    & 337 & 224 & 66.47 \\

Ordinary restoration of order
    & 302 & 32 & 10.60 \\

\midrule

Assignment contrast (percentage points)
    & 639 & -- & $+55.87^{***}$ $(3.13)$ \\

\bottomrule
\end{tabular}

\begin{tablenotes}
\footnotesize
\item \emph{Notes:}
Reanalysis of ICPSR 9982 police-record field recording
the person directed to leave.
The contrast uses original assignment, not a regression
on the order actually issued.
HC2 standard error in parentheses.
$^{*}p<0.10$, $^{**}p<0.05$, $^{***}p<0.01$.
A blank officer report is not proof that a person
remained in the residence; this outcome identifies
documented legal direction, not physical departure.
The usual-response arm can contain an order because
implemented responses need not coincide with
assigned protocols.
\end{tablenotes}

\end{threeparttable}
\end{table}
Table~\ref{tab:colorado-legal} makes explicit the distinction between policy assignment and implemented legal direction. Ordinary-response encounters can include an order, and an encounter assigned to an order need not produce a recorded direction to the suspect. Consequently, the effect is the difference in \emph{documented orders under randomized assignment}, not a comparison of officers who chose to issue an order. The outcome shows that the assignment materially alters the content of the police response: officers more frequently record a directive aimed at the suspected aggressor. Whether the direction results in movement is a separate behavioral question.

The legal-direction comparison is also an illustration of why the unit of treatment must remain visible. A police department can assign officers to use a response protocol, yet the response recorded in an individual case can reflect discretion, information at the scene, and incomplete documentation. Estimating the effect of the assignment preserves the randomization. Estimating an association between a recorded order and departure instead would compare cases selected by officers' implemented actions. The main police result retains the former contrast; the two observed outcomes---documented legal direction and scene movement---describe different ways in which the randomized intervention is expressed.

\subsection{The distribution of actual departures}
The pooled active-package contrast reduces victim-only departure by 11.97 percentage points and increases suspect-only departure by 9.43 points. The difference in the signed index is 21.40 points. In contrast, the estimated effect on \emph{any departure} is $-1.74$ points and is not statistically significant. This is the main empirical distinction: the observed distribution of who leaves changes substantially, while the indicator for any movement gives little indication of the reallocation.

\begin{table}[htbp]
\centering
\caption{Colorado: experimental response packages and the incidence of movement}
\label{tab:colorado-contrasts}
\begin{threeparttable}
\small
\begin{tabular}{p{5.0cm}rrr}
\toprule
Assignment versus ordinary response
    & Incidence index (pp)
    & Any departure (pp)
    & Observed T/C \\
\midrule

Arrest + EPO
    & $+24.08^{***}$
    & $-3.25$
    & 277 / 255 \\
    & $(5.10)$
    & $(4.31)$
    & \\

\addlinespace[0.5em]

Counseling + EPO
    & $+16.46^{***}$
    & $-5.77$
    & 282 / 255 \\
    & $(5.07)$
    & $(4.28)$
    & \\

\addlinespace[0.5em]

EPO alone
    & $+23.79^{***}$
    & $+3.96$
    & 273 / 255 \\
    & $(5.34)$
    & $(4.35)$
    & \\

\midrule

All active packages
    & $+21.40^{***}$
    & $-1.74$
    & 832 / 255 \\
    & $(4.41)$
    & $(3.57)$
    & \\

\bottomrule
\end{tabular}

\begin{tablenotes}
\footnotesize
\item \emph{Notes:}
Original-assignment differences relative to ordinary
response among cases with observed victim and suspect
departures. Signed incidence is defined as suspect-only
minus victim-only departure. HC2 standard errors are
reported in parentheses.
$^{*}p<0.10$, $^{**}p<0.05$,
$^{***}p<0.01$.
No outcomes are imputed for absent records.
Each active arm represents a distinct response package.
The pooled comparison combines all three active
assignments against ordinary response.
\end{tablenotes}
\end{threeparttable}
\end{table}

The signed index rises under each active package when compared with ordinary response: 24.08 points for arrest plus an order, 16.46 for counseling plus an order, and 23.79 for an order alone. These are effects of the packages as assigned. They do not isolate the incremental contributions of arrest, counseling, and the order as individual components. The arm-specific comparisons instead establish that the change in departure composition is visible across different kinds of active police response.

\begin{figure}[htbp]\centering
\includegraphics[width=.85\linewidth]{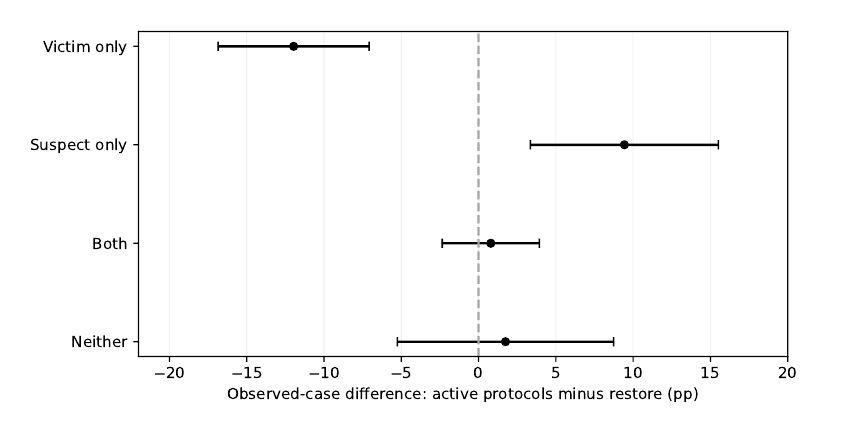}
\caption{Colorado: differences across the four exclusive scene states, pooled active protocols versus ordinary response. Horizontal intervals refer to the 1,087 cases with both departure indicators available.}\label{fig:colorado}\end{figure}

Figure~\ref{fig:colorado} shows the differences across the four exclusive departure states. The movement is concentrated in the two one-person categories. That pattern explains why the signed measure is more informative than the aggregate departure measure: it records the substitution between who leaves alone while preserving the both-and-neither categories in the full distribution.

A year-adjusted specification yields a 22.21-point index contrast, close to the unadjusted 21.40-point estimate. Among encounters involving female victims, the estimated contrast is 18.73 points. These comparisons use the encounters for which the corresponding indicators are observed. In the original administrative comparison, by contrast, the effect on the recorded suspect-directed order is estimated on all assignments in the order-only and ordinary-response arms.

I also examine later measures of safety that the Colorado experiment recorded. The pooled active-package estimate for an incident known to police within six months is $-1.00$ percentage points; for victim-reported physical assault within three months it is $-2.27$ points. Neither is statistically significant. These outcomes concern events after the intervention, with different collection processes. They should not be used interchangeably with departure at the encounter. The evidence from Colorado is most directly informative about the allocation of immediate movement and the legal directions that accompany the assigned response.

\section{Advocacy, detention, and private protective action}\label{sec:support}
\subsection{Advocacy and safety preparations}
Police orders and housing assistance do not exhaust the ways institutions can affect protective adjustment. Advocacy operates through assistance with the steps a person may choose to take. In Harris County, the randomized advocacy intervention provided support for safety planning, while subsequent interviews recorded specific preparations. Such preparations are informative about the capacity to act: an intervention may induce more of them because it makes the action feasible, not because it imposes a new obligation.

The estimates in Table~\ref{tab:harris} are baseline-adjusted effects on the fraction of observed follow-up waves with each preparation. Advocacy increases reports of hidden emergency cash by 20.08 percentage points, spare keys by 26.73 points, and an emergency bag by 30.16 points. The estimated difference in receipt of a protection order is 4.00 points and is not statistically significant. The sample contains 150 randomized women, with 148--149 valid observations for the preparation measures.

\begin{table}[htbp]
\centering
\caption{Harris County: advocacy and private safety preparations}
\label{tab:harris}
\begin{threeparttable}
\small
\begin{tabular}{p{8.5cm}rr}
\toprule
Outcome & Effect (pp) & N \\
\midrule

Protection order obtained
    & $+4.00$ & 150 \\
    & $(8.19)$ & \\

\addlinespace[0.5em]

Fraction of follow-ups with hidden emergency money
    & $+20.08^{***}$ & 149 \\
    & $(5.67)$ & \\

\addlinespace[0.5em]

Fraction of follow-ups with spare keys
    & $+26.73^{***}$ & 148 \\
    & $(5.43)$ & \\

\addlinespace[0.5em]

Fraction of follow-ups with emergency bag
    & $+30.16^{***}$ & 149 \\
    & $(5.96)$ & \\

\bottomrule
\end{tabular}

\begin{tablenotes}
\footnotesize
\item \emph{Notes:}
Original randomized assignment: 75 advocacy and
75 control. Estimates are intention-to-treat effects.
HC2 standard errors are reported in parentheses.
$^{*}p<0.10$, $^{**}p<0.05$,
$^{***}p<0.01$.
Each preparation outcome measures the fraction of
valid follow-up waves in which the corresponding
action was reported, adjusting for its baseline indicator.
Source response code 66 denotes not applicable
and is treated as missing, never as a negative response.
The advocacy intervention and related safety-planning
effects were previously studied in the original evaluation.
\end{tablenotes}
\end{threeparttable}
\end{table}

These effects are consistent with the previously documented safety-planning response to advocacy \citep{mcfarlane2002,mcfarlane2004}. Their role in this paper is conceptual and empirical: they illustrate that a larger observed supply of private precaution need not mean a larger externally imposed burden. Unlike an officer directing someone to leave, advocacy may provide resources or coordination that enable a person to carry out a precaution she already values. The three preparation measures therefore belong alongside, but not inside, the departure index used for Colorado.

\subsection{Detention and temporary relocation}
The Milwaukee experiment records a different margin after a police encounter: whether a person stayed elsewhere to avoid being hurt. Of 1,200 assigned cases, 919 have a valid answer to this item. Compared with advice, assignment to overnight arrest changes the probability of a temporary safety-motivated move by $-2.48$ points; short-release arrest changes it by 2.02 points. Neither difference is statistically significant.

\begin{table}[htbp]
\centering
\caption{Milwaukee: detention assignments and later outcomes}
\label{tab:milwaukee}
\begin{threeparttable}
\small
\begin{tabular}{p{8.5cm}rr}
\toprule
Contrast versus advice & Effect (pp) & N \\
\midrule

Overnight arrest: temporary move to avoid harm
    & $-2.48$ & 622 \\
    & $(3.44)$ & \\

\addlinespace[0.5em]

Short-release arrest: temporary move to avoid harm
    & $+2.02$ & 594 \\
    & $(3.62)$ & \\

\addlinespace[0.5em]

Overnight arrest: police-known same-pair repeat
    & $-0.24$ & 806 \\
    & $(3.23)$ & \\

\addlinespace[0.5em]

Short-release arrest: police-known same-pair repeat
    & $+0.56$ & 792 \\
    & $(3.27)$ & \\

\bottomrule
\end{tabular}

\begin{tablenotes}
\footnotesize
\item \emph{Notes:}
Original randomized experiment: 408 overnight arrest,
394 short-release arrest, and 398 advice assignments.
Estimates are intention-to-treat differences relative
to advice. Heteroskedasticity-robust standard errors
are reported in parentheses.
$^{*}p<0.10$, $^{**}p<0.05$,
$^{***}p<0.01$.
The temporary-move outcome (D34) is observed for
919 of the 1,200 original assignments.
Administrative repeat outcomes capture police-known
incidents involving the same pair, not the full
incidence of victimization.
\end{tablenotes}
\end{threeparttable}
\end{table}

The corresponding estimates for later police-known incidents involving the same pair are $-0.24$ and $+0.56$ points, and are also not statistically significant. Milwaukee's relocation item identifies the respondent's stated reason for a subsequent move: avoiding harm. That makes it different from the immediate movement recorded in Colorado and from relationship separation in the housing study. The estimates do not identify a change in later relocation attributable to the two arrest assignments relative to advice, but they show how a purpose-specific movement outcome can be incorporated into the evaluation of legal response.

Taken together, the supporting experiments prevent an overly narrow definition of protective action. Some interventions impose a direction at a particular encounter; others assist preparation; still others may affect later safety-motivated moves. The common measurement issue is to identify \emph{which action occurs and to whom}, rather than to regard every precaution or change of residence as the same outcome.

The Harris and Milwaukee results also help define the reach of the incidence concept. Harris measures repeated preparations that people may choose after receiving assistance; Milwaukee measures a temporary move made specifically to avoid harm. They do not reproduce either the housing-study joint residential state or the Colorado scene-departure index. Their value is that they prevent the concept of burden from being reduced to one physical action. The legal and economic interpretation depends on the nature of the action, the availability of support, and whether the response is directed by an institution or chosen by the individual.

\section{Implications for legal and policy design}\label{sec:discussion}
The evidence distinguishes two ways law and policy can change the incidence of protection. One expands the alternatives available to the person making a decision. The other changes which person a public authority directs to adjust at a particular encounter. Both affect the distribution of protective action, but their underlying instruments are different: a housing offer makes separation feasible; an emergency order assigns a direction to leave. Evaluating them requires outcome measures matched to those instruments.

The housing analysis shows the value of measuring the circumstances accompanying a relationship change. The same observed separation can occur with stable independent housing or with renewed residential instability. Those states have different implications for the feasibility of independent living, even without assigning a monetary cost or assuming that every separation was motivated by violence. Because the original study recruited families from shelters rather than from a register of known victims, this measure also captures the broader importance of outside options for relationship decisions.

The police analysis shows why institutional directions and aggregate departures should be distinguished. Prior evaluations of arrest and enforcement have emphasized deterrence and subsequent violence \citep{sherman1984,pate1992,maxwell2002,iyengar2009}; the scene-level decomposition captures a separate dimension of the same public response. Recording that someone left the scene does not reveal who had to leave. My decomposition identifies the person making the immediate move; the administrative order measure identifies the person to whom police recorded a legal direction. Under the active response packages, the observed distribution shifts away from victim-only departure and toward suspect-only departure. That reallocative dimension is of direct interest in the design of enforcement responses, regardless of whether the number of encounters involving any departure changes.

Finally, the Harris results show how outside assistance can increase private preparations by making them feasible. More preparation need not be worse for the person taking the precaution; fewer departures need not be better if the person who remains faces danger. An empirical measure of the \emph{burden of protection} should therefore begin with the action, its institutional source, and the feasible alternatives, rather than applying a single welfare sign to every move or precaution. The central lesson is that conventional policy outcomes can be supplemented by measures that identify the distribution of adjustment across people and circumstances.

The broader implication for policy evaluation is methodological as well as substantive. If separation and residential continuity are measured separately, a researcher cannot determine from their two averages how frequently the same family both separates and maintains a stable home. If a police record notes only that someone left, it cannot identify who undertook the move. These omissions are not merely matters of presentation. They change which aspects of a policy are visible, and therefore which legal and economic questions can be answered using the resulting estimates.

For housing assistance, a joint relationship--residential measure indicates whether a policy expands an outside option that can be exercised under durable conditions. For police response, a four-state departure classification indicates which individual makes the immediate adjustment, while the order record identifies the assignment of a legal direction. For advocacy, repeated preparations reveal an expansion of the actions an individual can undertake with support. None of these requires converting every behavioral response into a common monetary cost. The relevant incidence is recorded in the distribution of actions and opportunities itself.

\section{Conclusion}\label{sec:conclusion}
Protective institutions affect more than the probability of harm. They also shape the alternatives available to people at risk and the identity of those who must adjust when intervention occurs. I study these margins using randomized housing assistance and police responses in the United States.

In the housing experiment, the offer of potentially permanent assistance increases separation accompanied by stable independent residence. The result distinguishes a durable outside option from a separation followed by housing instability. In Colorado Springs, active police-response packages change who is observed leaving the scene even though the probability of any departure does not change significantly; assignment to an emergency order also increases documented directions for suspects to leave. Advocacy evidence shows that institutional assistance can expand private safety preparation.

These findings make the incidence of protective adjustment an observable outcome of policy evaluation. The relevant question is not simply whether a relationship ends or someone leaves. It is which alternatives become feasible and whose circumstances change as protection is put into practice. Recording those distinctions provides a fuller account of what legal and public interventions do.

\section*{Appendix}
\section*{Specification and measurement checks}\label{sec:robustness}
\subsection*{Housing comparisons across weights and survey waves}
The joint housing result is robust to the ordinary specification choices available in the public-use file. I hold fixed the original assignment comparison and the baseline-partnered population, changing whether the estimation is weighted and whether it controls for program availability observed before randomization. At 37 months, the raw unweighted difference in separation with stable independent housing is 12.82 percentage points; the availability-adjusted unweighted difference is 13.27; and the preferred weighted, adjusted difference is 12.52. Table~\ref{tab:hud-robust} reports the six contrasts.

\begin{table}[htbp]
\centering
\caption{Housing outside options: sensitivity to specification and survey wave}
\label{tab:hud-robust}
\begin{threeparttable}
\small
\begin{tabular}{p{8.5cm}rr}
\toprule
Specification for separation with stable housing
    & Effect (pp) & Observed N \\
\midrule

\multicolumn{3}{l}{\textit{37 months}} \\
\addlinespace[0.3em]

Unweighted, difference in means
    & $+12.82^{**}$ & 240 \\
    & $(5.12)$ & \\

\addlinespace[0.5em]

Unweighted, availability adjusted
    & $+13.27^{**}$ & 240 \\
    & $(5.21)$ & \\

\addlinespace[0.5em]

Weighted, availability adjusted
    & $+12.52^{**}$ & 240 \\
    & $(5.33)$ & \\

\bottomrule
\end{tabular}

\begin{tablenotes}
\footnotesize
\item \emph{Notes:}
All estimates compare assignment to permanent housing
subsidy with usual care among baseline-partnered
families with observed joint outcomes.
HC2 standard errors are reported in parentheses.
$^{*}p<0.10$, $^{**}p<0.05$,
$^{***}p<0.01$.
Weights are public-use wave- and comparison-specific
nonresponse weights. Program availability is measured
before randomization. The preferred specification
is weighted and availability adjusted.
\end{tablenotes}

\end{threeparttable}
\end{table}

The stability of the 37-month estimates across these specifications shows that the reported joint effect is not an artifact of adding the pre-assignment availability controls or using the comparison-specific weight. 

The joint measure also differs from a regression that first selects on separation and then asks who became stably housed. Separation is itself affected by housing assistance; selecting on it would redefine the treatment and comparison groups after assignment. Cross-classifying the original relationship and residential indicators avoids that conditioning. The analysis thus provides the experimental distribution of \emph{combined} states among the available linked follow-up responses.

All three 37-month estimates use the same joint-state definition and the same eligible initially partnered comparison; the specifications change only the weighting convention and whether the pre-treatment assignment-set variables enter. The main coefficient remains close to 12--13 percentage points. 

\subsection*{Police response packages and recorded outcomes}
In Colorado, the positive change in the signed incidence index appears under each of the three separately assigned active protocols. Adding year effects changes the pooled estimate from 21.40 to 22.21 points. A female-victim subsample analysis yields 18.73 points. These checks concern the composition of observed departures and use the original treatment assignment throughout.

A separate check uses the officer's recorded direction to leave rather than the scene-departure indicators. Because it is available for the full order-only versus ordinary-response comparison, it directly captures one institutional component of the policy and does not require both scene indicators. Its 55.87-point effect is not numerically comparable with the signed departure index: one is an indicator for a recorded instruction, the other a contrast between two kinds of movement. Considering them together establishes that legal direction and actual behavior are separately observable parts of the response.

The two kinds of Colorado outcome also have different evidentiary strengths. The police-recorded legal direction is available for the entire original assignment comparison between order-only and ordinary response. The paired scene outcomes permit a richer behavioral classification, but only where both reports are present. The full-cohort legal result and observed-case movement result are therefore mutually informative rather than alternative estimates of one effect. The first establishes a substantial shift in documented institutional direction; the second shows a corresponding change in the observed composition of immediate departure across response packages.

\bibliographystyle{apalike}

\begin{thebibliography}{99}
\bibitem[Adams et al.(2024)]{adams2024} Adams, Abi, Kristiina Huttunen, Emily Nix, and Ning Zhang (2024). ``The Dynamics of Abusive Relationships.'' \emph{Quarterly Journal of Economics} 139(4), 2135--2180. \url{https://doi.org/10.1093/qje/qjae022}.
\bibitem[Aizer(2010)]{aizer2010} Aizer, Anna (2010). ``The Gender Wage Gap and Domestic Violence.'' \emph{American Economic Review} 100(4), 1847--1859. \url{https://doi.org/10.1257/aer.100.4.1847}.
\bibitem[Berk et al.(1992)]{berk1992} Berk, Richard A., Alec Campbell, Ruth Klap, and Bruce Western (1992). ``A Bayesian Analysis of the Colorado Springs Spouse Abuse Experiment.'' \emph{Journal of Criminal Law and Criminology} 83(1), 170--200. \url{https://doi.org/10.2307/1143828}.
\bibitem[Black et al.(2006)]{icpsr9982} Black, Howard, Richard A. Berk, James Lily, Robert Owenbey, and Giannina Rikoski (2006). \emph{Evaluating Alternative Police Responses to Spouse Assault in Colorado Springs: An Enhanced Replication of the Minneapolis Experiment, 1987--1989}. ICPSR 9982, version 1. Ann Arbor, MI: Inter-university Consortium for Political and Social Research. \url{https://doi.org/10.3886/ICPSR09982.v1}.
\bibitem[Bloch and Rao(2002)]{bloch2002} Bloch, Francis, and Vijayendra Rao (2002). ``Terror as a Bargaining Instrument: A Case Study of Dowry Violence in Rural India.'' \emph{American Economic Review} 92(4), 1029--1043. \url{https://doi.org/10.1257/00028280260344588}.
\bibitem[Bobonis et al.(2013)]{bobonis2013} Bobonis, Gustavo J., Melissa Gonz\'alez-Brenes, and Roberto Castro (2013). ``Public Transfers and Domestic Violence: The Roles of Private Information and Spousal Control.'' \emph{American Economic Journal: Economic Policy} 5(1), 179--205. \url{https://doi.org/10.1257/pol.5.1.179}.
\bibitem[Browne and Bassuk(1997)]{browne1997} Browne, Angela, and Shari S. Bassuk (1997). ``Intimate Violence in the Lives of Homeless and Poor Housed Women: Prevalence and Patterns in an Ethnically Diverse Sample.'' \emph{American Journal of Orthopsychiatry} 67(2), 261--278. \url{https://doi.org/10.1037/h0080230}.
\bibitem[Dee(2003)]{dee2003} Dee, Thomas S. (2003). ``Until Death Do You Part: The Effects of Unilateral Divorce on Spousal Homicides.'' \emph{Economic Inquiry} 41(1), 163--182. \url{https://doi.org/10.1093/ei/41.1.163}.
\bibitem[Desmond and Kimbro(2015)]{desmond2015} Desmond, Matthew, and Rachel Tolbert Kimbro (2015). ``Eviction's Fallout: Housing, Hardship, and Health.'' \emph{Social Forces} 94(1), 295--324. \url{https://doi.org/10.1093/sf/sov044}.
\bibitem[Dugan et al.(2003)]{dugan2003} Dugan, Laura, Daniel S. Nagin, and Richard Rosenfeld (2003). ``Exposure Reduction or Retaliation? The Effects of Domestic Violence Resources on Intimate-Partner Homicide.'' \emph{Law \& Society Review} 37(1), 169--198. \url{https://doi.org/10.1111/1540-5893.3701005}.
\bibitem[Ehrlich and Becker(1972)]{ehrlich1972} Ehrlich, Isaac, and Gary S. Becker (1972). ``Market Insurance, Self-Insurance, and Self-Protection.'' \emph{Journal of Political Economy} 80(4), 623--648. \url{https://doi.org/10.1086/259916}.
\bibitem[Eswaran and Malhotra(2011)]{eswaran2011} Eswaran, Mukesh, and Nisha Malhotra (2011). ``Domestic Violence and Women's Autonomy in Developing Countries: Theory and Evidence.'' \emph{Canadian Journal of Economics} 44(4), 1222--1263. \url{https://doi.org/10.1111/j.1540-5982.2011.01673.x}.
\bibitem[Farmer and Tiefenthaler(1997)]{farmer1997} Farmer, Amy, and Jill Tiefenthaler (1997). ``An Economic Analysis of Domestic Violence.'' \emph{Review of Social Economy} 55(3), 337--358. \url{https://doi.org/10.1080/00346769700000004}.
\bibitem[Gubits et al.(2016)]{gubits2016} Gubits, Daniel, Marybeth Shinn, Michelle Wood, Stephen Bell, Samuel Dastrup, Claudia Solari, Scott Brown, Debi McInnis, Tom McCall, and Utsav Kattel (2016). \emph{Family Options Study: 3-Year Impacts of Housing and Services Interventions for Homeless Families}. Washington, DC: U.S. Department of Housing and Urban Development. \url{https://www.huduser.gov/portal/sites/default/files/pdf/Family-Options-Study-Full-Report.pdf}.
\bibitem[Iyengar(2009)]{iyengar2009} Iyengar, Radha (2009). ``Does the Certainty of Arrest Reduce Domestic Violence? Evidence from Mandatory and Recommended Arrest Laws.'' \emph{Journal of Public Economics} 93(1--2), 85--98. \url{https://doi.org/10.1016/j.jpubeco.2008.09.006}.
\bibitem[Maxwell et al.(2002)]{maxwell2002} Maxwell, Christopher D., Joel H. Garner, and Jeffrey A. Fagan (2002). ``The Preventive Effects of Arrest on Intimate Partner Violence: Research, Policy and Theory.'' \emph{Criminology \& Public Policy} 2(1), 51--80. \url{https://doi.org/10.1111/j.1745-9133.2002.tb00107.x}.
\bibitem[McFarlane and Malecha(2003)]{icpsr3542} McFarlane, Judith, and Ann Malecha (2003). \emph{Evaluating a Collaborative Intervention Between Health Care and Criminal Justice in Harris County, Texas, 2001--2002}. ICPSR 3542, version 1. Ann Arbor, MI: Inter-university Consortium for Political and Social Research. \url{https://doi.org/10.3886/ICPSR03542.v1}.
\bibitem[McFarlane et al.(2002)]{mcfarlane2002} McFarlane, Judith, Ann Malecha, Julia Gist, Kathy Watson, Elizabeth Batten, Iva Hall, and Sheila Smith (2002). ``An Intervention to Increase Safety Behaviors of Abused Women: Results of a Randomized Clinical Trial.'' \emph{Nursing Research} 51(6), 347--354.
\bibitem[McFarlane et al.(2004)]{mcfarlane2004} McFarlane, Judith, Ann Malecha, Julia Gist, Kathy Watson, Elizabeth Batten, Iva Hall, and Sheila Smith (2004). ``Increasing the Safety-Promoting Behaviors of Abused Women.'' \emph{American Journal of Nursing} 104(3), 40--50.
\bibitem[Pate and Hamilton(1992)]{pate1992} Pate, Anthony M., and Edwin E. Hamilton (1992). ``Formal and Informal Deterrents to Domestic Violence: The Dade County Spouse Assault Experiment.'' \emph{American Sociological Review} 57(5), 691--697.
\bibitem[Shavell(1991)]{shavell1991} Shavell, Steven (1991). ``Individual Precautions to Prevent Theft: Private versus Socially Optimal Behavior.'' \emph{International Review of Law and Economics} 11(2), 123--132.
\bibitem[Sherman and Berk(1984)]{sherman1984} Sherman, Lawrence W., and Richard A. Berk (1984). ``The Specific Deterrent Effects of Arrest for Domestic Assault.'' \emph{American Sociological Review} 49(2), 261--272.
\bibitem[Sherman et al.(1992)]{sherman1992} Sherman, Lawrence W., Janell D. Schmidt, Dennis P. Rogan, Douglas A. Smith, Patrick R. Gartin, Ellen G. Cohn, Dean J. Collins, and Anthony R. Bacich (1992). ``The Variable Effects of Arrest on Criminal Careers: The Milwaukee Domestic Violence Experiment.'' \emph{Journal of Criminal Law and Criminology} 83(1), 137--169.
\bibitem[Sherman et al.(1994)]{icpsr9966} Sherman, Lawrence W., Janell D. Schmidt, and Dennis P. Rogan (1994). \emph{Milwaukee Domestic Violence Experiment, 1987--1989}. ICPSR 9966, version 1. Ann Arbor, MI: Inter-university Consortium for Political and Social Research. \url{https://doi.org/10.3886/ICPSR09966.v1}.
\bibitem[Stevenson and Wolfers(2006)]{stevenson2006} Stevenson, Betsey, and Justin Wolfers (2006). ``Bargaining in the Shadow of the Law: Divorce Laws and Family Distress.'' \emph{Quarterly Journal of Economics} 121(1), 267--288. \url{https://doi.org/10.1093/qje/121.1.267}.
\bibitem[Tauchen et al.(1991)]{tauchen1991} Tauchen, Helen V., Ann Dryden Witte, and Sharon K. Long (1991). ``Domestic Violence: A Nonrandom Affair.'' \emph{International Economic Review} 32(2), 491--511.
\bibitem[Winichakul and Zhang(2026)]{winichakul2026} Winichakul, K. Pun, and Ning Zhang (2026). ``The Effect of Housing Access on Intimate Partner Violence: Evidence from the Family Options Study.'' RFBerlin Discussion Paper 26105. \url{https://www.rfberlin.com/network-paper/the-effect-of-housing-access-on-intimate-partner-violence-evidence-from-the-family-options-study/}.
\end{thebibliography}

\end{document}